# Modular Reactivation of Mexico City After COVID-19 Lockdown

Guillermo de Anda-Jáuregui [1,2,3,*], PhD; Enrique Hernández-Lemus, PhD [1,2]
1. Computational Genomics Division, National Institute of Genomic Medicine, México City, México
2. Center for Complexity Sciences, Universidad Nacional Autónoma de México, México City Mexico
3. Programa de Cátedras CONACYT, Consejo Nacional de Ciencia y Tecnología, México City, México
* Corresponding author: gdeanda@inmegen.edu.mx

## Abstract

During the COVID-19 pandemic, the slope of the epidemic curve in Mexico City has been quite unstable. We have predicted that in the case that a fraction of the population above a certain threshold returns to the public space, the negative tendency of the epidemic curve will revert. Such predictions were based on modeling the reactivation of economic activity after lockdown by means of an epidemiological model resting upon a contact network of Mexico City derived from mobile device co-localization. We evaluated the epidemic dynamics considering the tally of active and recovered cases documented in the mexican government's open database. Scenarios were modeled in which different percentages of the population are reintegrated to the public space by scanning values ranging from 5 % (around 350,000; close to the number of people estimated by the local authorities to be returning to the public space) up to 50 %. Null models were built by using data from the Jornada Nacional de Sana Distancia (the Mexican model of elective lockdown) in which there was a mobility reduction of 75 % and no mandatory mobility restrictions. We found that a new peak of cases in the epidemic curve was very likely for scenarios in which more than 5% of the population rejoined the public space; The return of more than 50 % of the population synchronously will unleash a peak of a magnitude similar to the one that was predicted with no mitigation strategies. By evaluating the tendencies of the epidemic dynamics, the number of new cases registered, new cases hospitalized, and new deaths, we consider that under this scenario, reactivation following only elective measures may not be optimal. Given the need to reactivate economic activities, we suggest to consider alternative measures that allow to diminish the contacts among people returning to the public space. We evaluated that by "encapsulating" reactivated workers (that is, using measures to reduce the number of contacts beyond their effective community in the contact network) may allow a reactivation of a larger fraction of the population without compromising the desired tendency in the epidemic curve.

# 1. Introduction

COVID-19, the disease caused by the novel coronavirus SARS-CoV-2 is a complex pathology of infectious origins. The pathology is contagious via the airway (very likely even air-borne) [Stadnytskyi, et al 2020] and is able to unleash, not only the characteristic severe acute respiratory syndrome (SARS) both also a series of clinical manifestations of immune and inflammatory type that may lead to pneumonia and even lead to sepsis induced systemic failure (SISF) [Zaim, et al, 2020]. Being a nascent infectious epidemic, we have little information regarding its broad clinical manifestations, even less regarding the molecular mechanisms behind. However, it is known that the disease is able to adopt a broad range of forms, from a mild, temporary respiratory infectious disease to a complex pathology with a high mortality rate, often requiring critical care.

The enormous burden of disease caused by SARS-CoV2 has been a consequence of the highly intertwined chains of transmission of the epidemic over extensively interconnected contact networks, local, regional and global [Yamamoto, et al 2020]. For this reason, facing the challenges posed by this disease involves understanding the way dynamic epidemic processes happen in complex human interaction networks [Altohouse, et al 2020; Kojaku, et al 2020].

## 1.1 A primer on network epidemiology

Network Epidemiology or Epidemics on Networks (EoN) has been defined as the study of the spread of disease and risky behaviors among populations founded in the tenets of network science [Eames and Read, 2007; Kiss, et al 2017].

EoN is then concerned on the modelisation of disease spreading, as well as contagion and diffusion processes happening amidst social (especially public) spaces in living systems. Tha main emphasis to date has been applied to human populations, however similar methods can be easily adapted to deal with animal plagues, epidemics on livestock and so on. The aim of EoN is being able to build realistic, mechanistic models to explain the spread of human disease by considering individual and collective mobility as well as population and meta-population features influencing the contact between individuals leading to contagion events.

These models may in turn be used to forecast the spreading of diseases (infectious and otherwise) by agents. Such agents may be infectious organisms such as bacteria, virus, and the likes, both also social behaviors such as the propensity to smoke, overeating and other factors influencing chronic diseases with a collective dimension component. As in clinical and social epidemiology, the goal is to be able to determine risks, containment strategies and to be able to assess targeted interventions [Kiss, et al 2017].

The usual setting of EoN considers the individuals in a population as nodes in a network. The interactions connecting the individuals (i.e. the links) are such that epidemic behavior emerges from these, such as contacts leading to the propagation of contagious agents (pathogens). The structure of the network influences the dynamic behavior of the epidemic and may offer clues as to what kind of mitigation and containment strategies can be used.

## 2. Materials and Methods

### 2.1 Network modelling of human interactions using mobile devices:

Human contact networks are characterized by an heterogeneous connectivity degree distribution with a long tail. The properties of this type of networks are the object of study of the discipline called Network Science [Albert & Barabasi, 2001]. Such heterogeneous connectivity patterns have some consequences in the epidemiological setting such as a high variance in the individual reproductive number and the dominance of the superspreading events that arise due to the high degree nodes [Kiss, et al 2017].

A major challenge for EoN is the accurate representation of contact networks for a given population. Using a combination of observational and technology-based methods, it is possible to reconstruct contact networks in limited, well-bound settings such as schools [Stehlé et al, 2011], hospitals [Isella *et al* 2011], or conferences [Szomszor *et al* 2010]. However, the challenge of reconstructing a contact network at the scale of a city is non-trivial. Recently, a contact network for Mexico City was reconstructed through the use of anonymized mobile device locations throughout a single day **[de-Anda Jáuregui et al 2020]**. This network was released as open data [doi:10.17605/OSF.IO/B6G92].

In order to be able to effectively run epidemic simulations, we scaled down the large contact network for Mexico City using a methodology based on the stochastic blockmodel (SBM) structure of the original network [Peixoto 2014, Peixoto 2015]. Briefly, we obtained the SBM structure of the largest connected component of the contact network of Mexico City, and scaled the size of each individual block to 1/10th of the original. Then, we generated a new network using the original SBM edge probability on the scaled down blocks. This new network captures the original's topology in terms of degree distribution and clustering coefficients, which is necessary for the results of an EoN dynamics to be representative of the larger one [Kiss, et al 2017]. These network operations were performed using the {graph-tool} package [Peixoto 2014b].

## 3. Theory/Calculations

### 3.1 Model Fundamentals and assumptions

In this study, we will be working under the following set of assumptions:

1. The disease in Mexico City is transmitted over a heterogeneous network reflecting the highly hierarchical and modular structure of urban systems [Levinson, 2012]

2. Epidemic dynamics is guided, in part, by factors intrinsic to the virus which exhibit stochastic behavior as a product of interindividual biological heterogeneity [Zhao and Chen, 2020].

3. In view of assumptions 1 and 2, the distribution of the number of contagions induced by each infected individual exhibits a long tail, giving rise to the existence of certain nodes acting as contagion hubs. This in turn led to the formation of super-spreading events.

Under these assumptions, we used the previously described contact network for Mexico City for our modelling purposes. A node in this contact network represents an inhabitant of Mexico City, and a link represents a close-range physical interaction between people; considering the resolution reported in the original manuscript (less than 2 m), these contacts can potentially transmit the infectious agent.

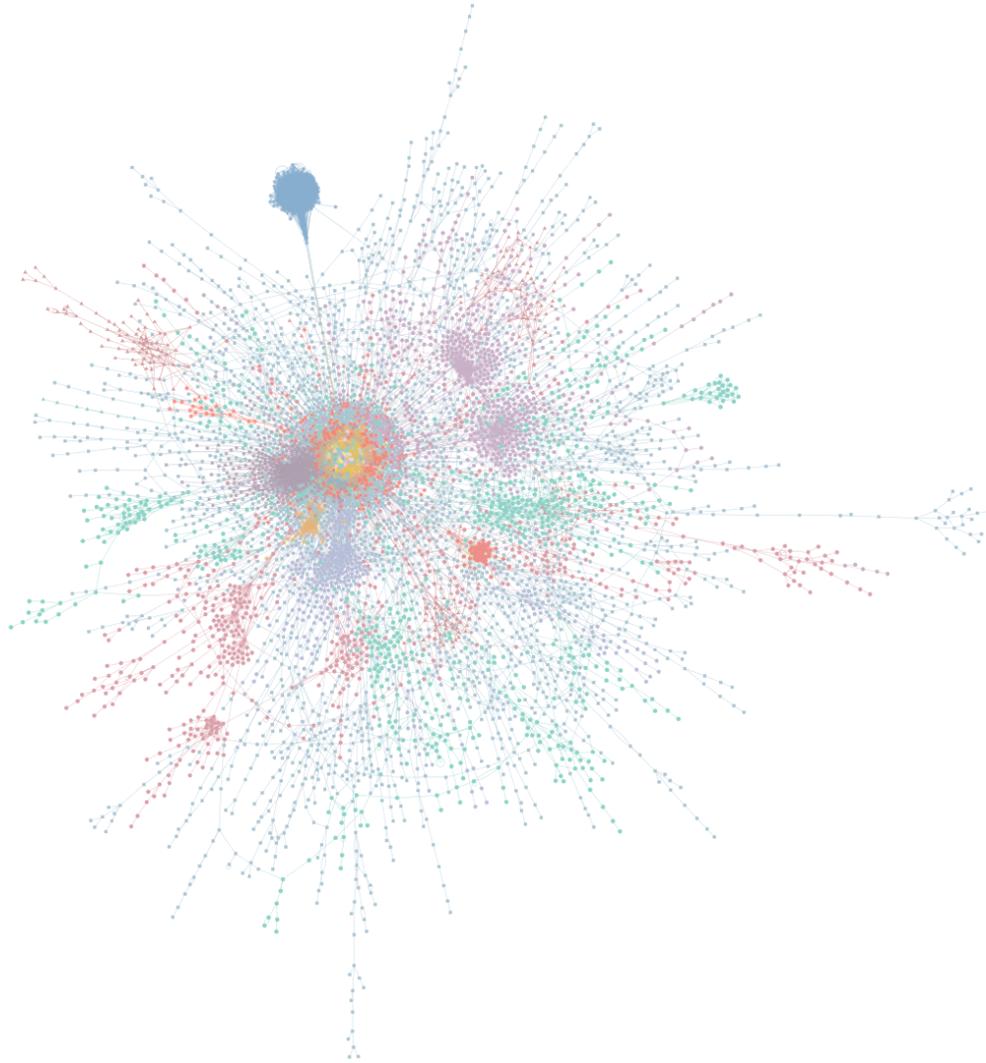

Figure 1: A network, with 7216 vertices and 86775 edges, representative of the close contact dynamics in Mexico City. The color of each node represents its "community," a module of nodes which are more closely connected to each other. We call this the CDMX network.

Once we have a somewhat reliable model for the human contact network structure of Mexico City, we are in a position to consider epidemic processes (namely, COVID-19 epidemics) happening on top of that network structure. The following subsection will be devoted to this.

### 3.2 Epidemiological simulation

We performed simulations of the epidemic dynamics by using the Epidemics on Network package (EoN, https://doi.org/10.21105/joss.01731) [Miller & Ting 2020] in order to simulate possible trajectories that the epidemic phenomena in different economic reactivation scenarios. To do so, we used a stochastic

Susceptible-Exposed-Infected-Recovered model on the Mexico City contact network.

To capture the interindividual variability in terms of incubation and recovery times, we used the parameters reported for the official Mexico City's government model (https://modelo.covid19.cdmx.gob.mx/) to generate a uniform distribution of values, such that each individual node has an incubation and recovery time between within half and twice the value used for the official model.

Since by the time period studied, general hygiene measures such as widespread use of face masks have been adopted, we decided to calculate the transmission rate for our model using the average Rt for the month of June, calculated as described in [Salas, 2020]. We used this value as the mean of a normal distribution, to account for the fact that certain contact behaviours are more conductive to disease transmission.

Epidemic dynamic calculations over realistic contact networks allow us for the modelisation of what may happen under different reactivation scenarios amidst the ongoing Covid-19 surges. In particular, this may open the way to educated predictions regarding risks for outbreaks of different magnitudes as we will see in the next subsection.

### 3.3 Risk assessment of reactivation schemes

The initial response to the COVID-19 emergency in Mexico was a voluntary lockdown known as *Jornada Nacional de Sana Distancia* (JNSD; literally: "Healthy Distance National Period"). During this period, non-essential economic activities were limited. Based on official figures [https://covid19.sinave.gob.mx/], there was a 75% mobility reduction within Mexico City. We represent the effect of such lockdown in network terms, by taking the original CDMX network and randomly removing 75% of its edge set; we refer to this as the JNSD network.

Any economic reactivation following the lockdown period implies that part of the city's population will return to its regular work activities. This will lead to a reemergence of contacts between people returning to the public space. In network terms, this is akin to the reemergence of links adjacent to the nodes representing people returning to the public space.

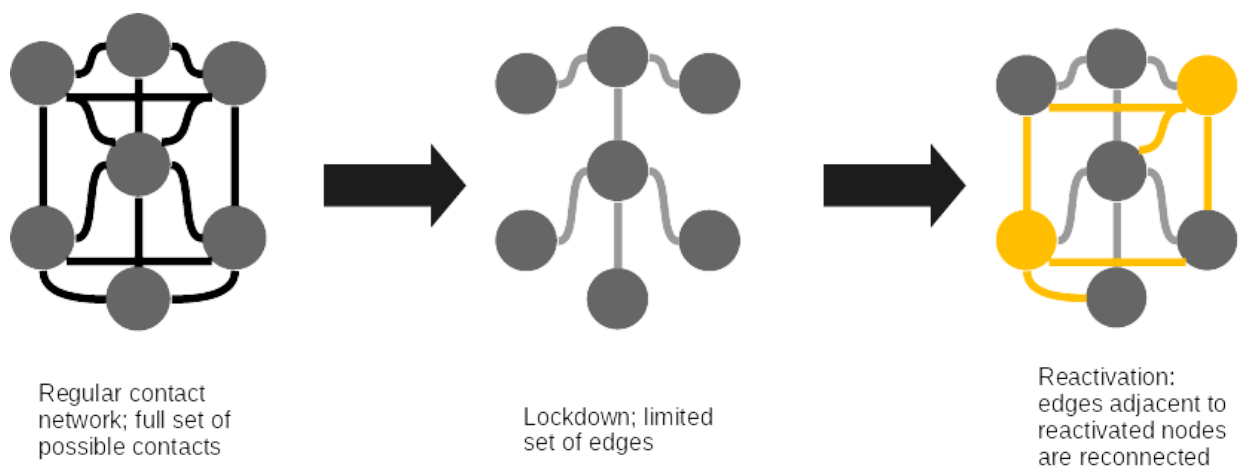

Figure 2: Schematic representation of the changes in contact networks induced by lockdowns and economic reactivation. Lockdowns induce disconnection of some links between nodes; whereas economic reactivation involves the reconnection of links adjacent to nodes representing people returning to the public space.

With this in mind, we can represent different scenarios of economic reactivation, by reconnecting nodes that were removed in the JNSD network, and simulating epidemic dynamics with the aforementioned parameters on them. For every scenario analysed, we run 100 iterations of EoN dynamics, and evaluate the behaviour in terms of:

i) whether a peak in the number of active infections occurs: measured as the number of simulations in which there is at least one day with more infected nodes than day 0.

ii) Percentage of infected population at the peak: measured as the average magnitude of the peak for all simulations.

iii) Peak time: measured as the average peak time for all simulations.

With these ideas in mind, we will present the epidemic risk evaluation under different reactivation scenarios.

### 3.3.1 Minimum and maximum contact scenarios - JNSD and CDMX networks

The range of scenarios is bound on one end by the minimum level of contact that happened during the lockdown, and is represented by the JNSD network. On the other side of the spectrum, the maximum level of contact is found in the CDMX network, which captures the usual contact patterns of the City without the constraints induced by the pandemic.

### 3.3.2 Scenarios of constraint-free reactivation

As we previously mentioned, economic reactivation involves the reintegration of a fraction of the population back into the public space. This renewed activity can be represented as a reconnection of those contacts that were removed during the lockdown.

Without any constraint and no additional information, it can be assumed that the workforce is evenly distributed within the population represented in the contact network [Freire 2010]. Therefore, the reactivation of any fraction of the population will be akin to randomly sampling the nodes of the contact network, and reconnecting the links lost during the lockdown. This is a conservative assumption due to lack of further information on the sociodemographic and spatial distribution of the individuals involved in the activities that will be reactivated.

We evaluate this scenario for different fractions of population ranging from 5% to 50%, using the following algorithm.

For each fraction of reactivated population F:

1) We selected F nodes of the Mexico City contact network
2) We considered that all links to nodes adjacent to these nodes are unrestrictedly re-activated.
3) We considered that the rest of the nodes in F (those with no correspondence to links in F) are the ones in the network as modeled under lockdown conditions as given by JNSD network.

We used the resulting network to run an EoN dynamic using the parameters described previously.

### 3.3.3 Scenarios of modular reactivation

An alternative approach to economic reactivation consists of leveraging network properties in order to impose limits to the epidemic dynamic. The concept of modularity in complex networks [Girvan & Newman, 2004] A module in a complex network is loosely defined as a set of nodes (individuals) with a higher number of connections among members of the set than with other nodes of the network, i.e. there are more connections within a module than between modules. An important property of modular networks is that dynamic phenomena (such as a random walk or pathogen propagation) tends to remain inside a module for a longer time before spreading outside the module.

In our proposed modular reactivation, instead of reactivating nodes spread throughout the network, whole modules are reactivated, until the desired fraction of the population is reactivated. All edges adjacent to the reactivated nodes are reconnected as well; it can be proven that this will include all links within a given module, and a smaller subset of links beyond the boundaries of the module.

In **Figure 3** we illustrate and contrast a constraint-free and a modular reactivation strategy.

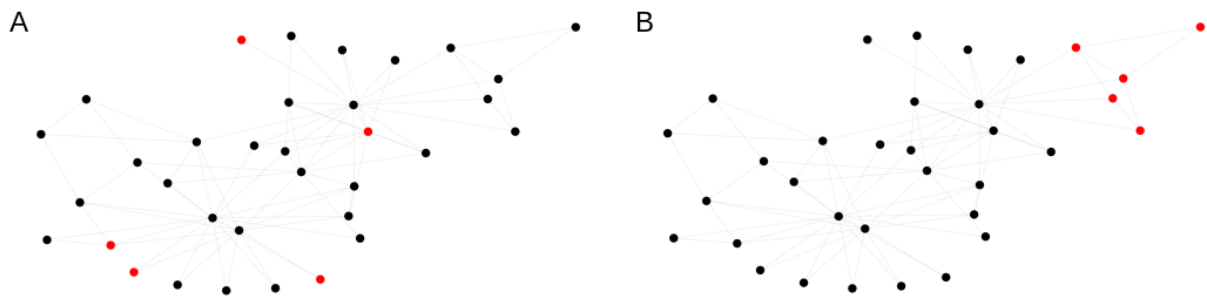

*Figure 3. Modular interconnections and contact structure. Panel A: constraint-free reactivation. Panel B: modular reactivation. Nodes to be reactivated are highlighted in red; the rest are shown in black.*

In panel A, the contacts of the red nodes may include job and non-job related contacts with similar probabilities. This way, by being randomly distributed in the network, contacts are reactivated both, with other essential workers (red) and with the rest of the population (black nodes) which in principle must remain under confinement (as during the lockdown).

Panel B, depicts a scenario in which red nodes form a module, hence, it is more likely that they connect to other essential workers (red nodes). In case of contagion within the module, the general population is, to an extent, shielded since the outbreak has a greater probability to keep spreading within the module. Hence the outbreak will become contained more easily.

We evaluated epidemic dynamics under these modularization scenarios, as presented in Figure 1 in the intermediate panels. We can see that allowing modularized contact network structures by encapsulating essential workers, the percentage of the population that may return to public space activities without causing major outbreaks is significantly higher, up to 25 % with a "smart modularized" strategy.

It is however relevant to highlight that such strategies may require strict adherence to the confinement scheme. This may imply the implementation of measures such as relocation of essential workers to housing near the workspaces, dedicated transportation and so on.

### 3.3.4 Leveraging intermodule connectivity to optimize modular reactivation

While modular reactivation can be accomplished by arbitrarily activating modules, the interconnection between modules can also be used to further refine the modular reactivation. Dynamic processes will tend to remain within a modular structure; therefore, reactivating smaller, topologically distant modules in the network will further encapsulate the epidemic phenomena and minimize its spread.

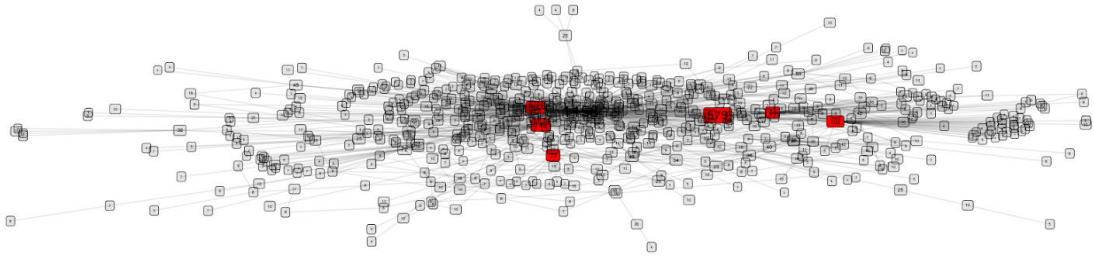

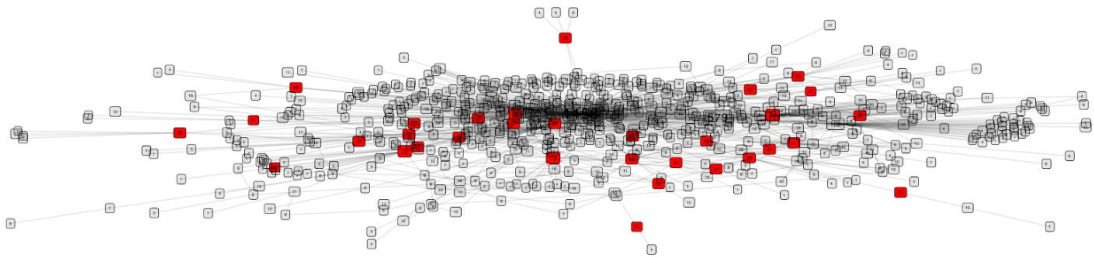

Figure 4: These are module projections of the CDMX contact network; each node represents a module in the contact network, with links representing intermodule connections (that is, contacts between nodes belonging to different modules). In both panels, the sum of nodes in the selected (red) modules contains roughly 20% of the network's population. In panel A, this fraction is concentrated in 6 modules, whereas in panel B it is spread over 33 smaller modules.

## 4. Results

In what follows we will present the results of several reactivation interventions in the predicted behavior of the epidemic curves in Mexico City. In order to comply with our scaling without imposing ecological fallacy biases, all our reports of epidemic curves will be expressed as a percentage of the population.

### 4.1 Constraint free reactivation quickly approaches the behaviour of a full reactivation

One initial, non-surprising result is that unconstrained reactivation of increasing fractions of the population, with no use of knowledge about the human contact network structure of Mexico City, rapidly approaches the epidemic conditions of full reactivation (Figure 5).

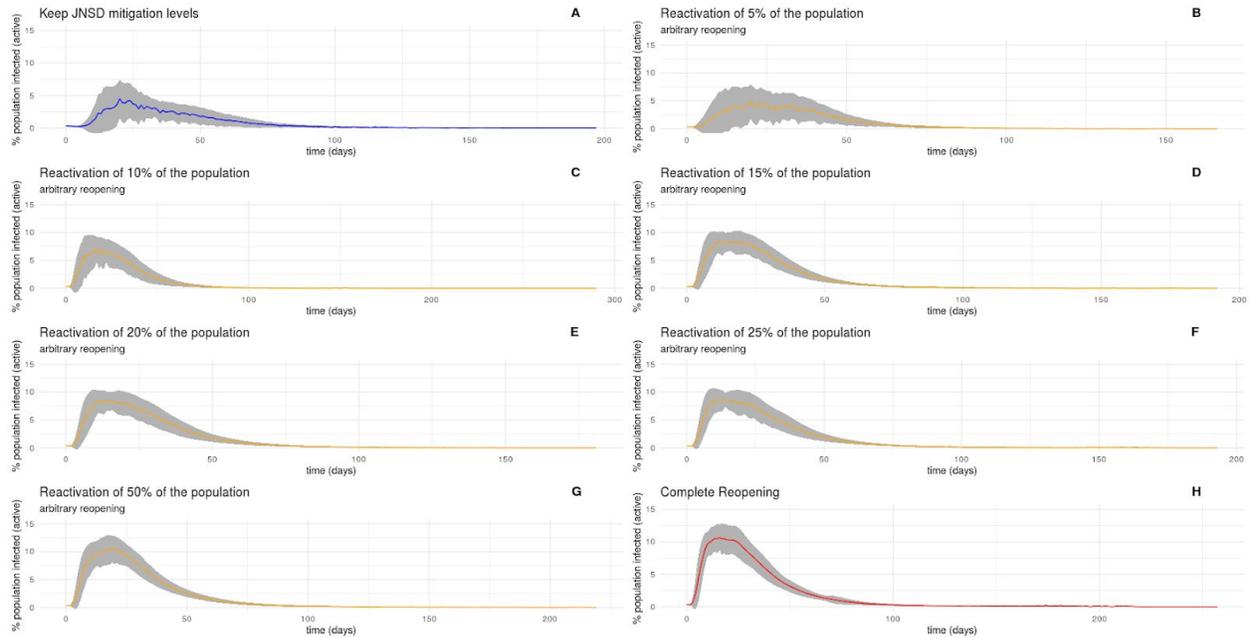

Figure 5: Ensemble visualization of epidemic simulations on networks. The colored line is traced on the average infected population percentage for each time point. The shadowed area represents ±1 standard deviation. Panel A is from the simulation using the JNSD network, representing a full lockdown; panels B through G represents dynamics on networks with a reactivation of 5% through 50% of the population, sampled without constraints from the network. Panel H represents the dynamics on a fully reactivated network.

Summarizing the results of the simulations, Table 1 shows that under different unconstrained scenarios, as the fraction of the population returning to the open public spaces increases, the behavior of a full reactivation (high peaks, longer growth curves and higher likelihood of outbreaks) is recovered.

Table 1: Summary statistics for constraint free reactivation strategies

| Reactivation scenario | Average Infection Peak Magnitude (%population) | Average Peak Time (days) | Peak likelihood |
|---|---|---|---|
| 0% (Lockdown continued as during JNSD) | 2.560421 | 5.436402 | 31% |
| 5% | 3.085227 | 4.668209 | 34% |
| 10% | 6.983370 | 12.705593 | 77% |
| 15& | 9.493764 | 15.403023 | 97% |
| 20% | 9.570953 | 14.997234 | 97% |
| 25% | 9.802384 | 15.311280 | 97% |
| 50% | 11.986835 | 18.733040 | 100% |
| 100% (full return to regular activity) | 12.282289 | 18.420751 | 100% |

Reactivation strategies in which the population is randomly distributed within the population allow for only a very small fraction of the population to be reactivated without a high risk for a resurgence of cases. We observe that any fraction beyond 5% of the population rejoining the public space leads to a high risk of a new peak to appear. Furthermore, a reactivation of 50% of the population is virtually indistinguishable to a full reactivation in terms of the magnitude, timing, and likelihood of a new peak.

## 4.2 Modular reactivation is better tolerated in terms of peak magnitude and time

A second, less dramatic scenario occurs whenever we make use of some knowledge about the human contact network structure of Mexico City, namely its modular character. The height and duration of the peaks as shown in Figure 6 is diminished (for similar percentages of returning population) with respect to the epidemic curves in the unconstrained case (Figure 5).

A quick glance at the summary statistics presented in Table 2 in comparison with Table 1 also reveals that this is actually the case.

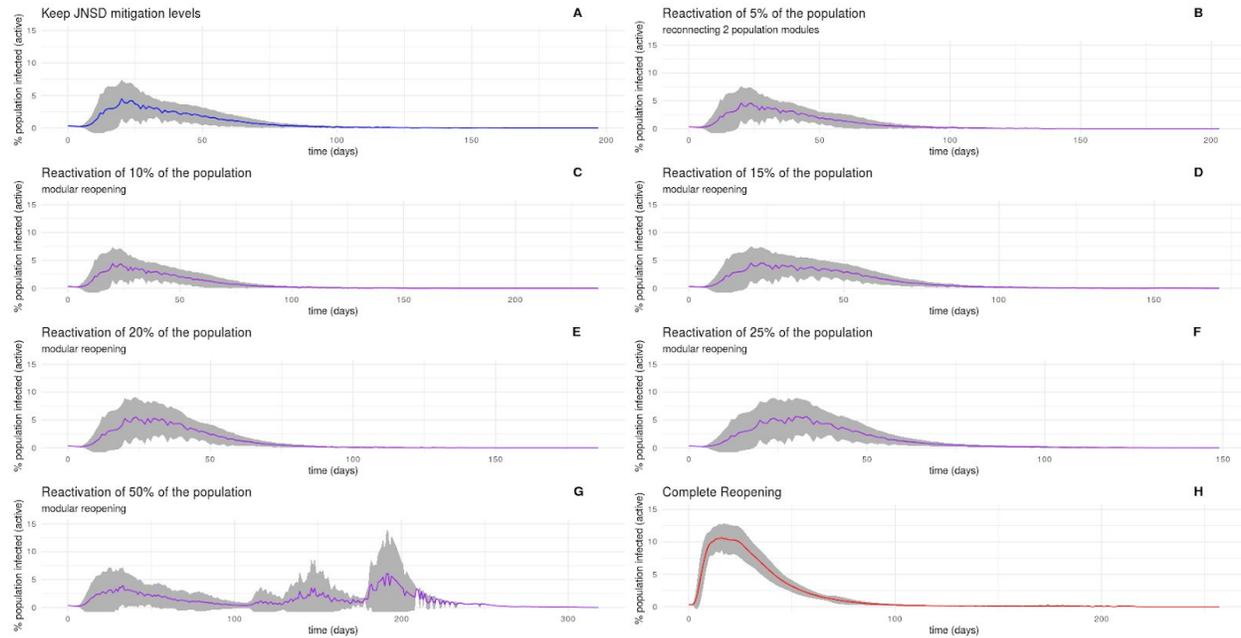

Figure 5: Ensemble visualization of epidemic simulations on networks. The colored line is traced on the average infected population percentage for each time point. The shadowed area represents ±1 standard deviation. Panel A is from the simulation using the JNSD network, representing a full lockdown; panels B through G represents dynamics on networks with a reactivation of 5% through 50% of the population, achieved through the reactivation of modules in the contact network. Panel H represents the dynamics on a fully reactivated network.

Table 2: Summary statistics of modular reactivation

| Reactivation scenario | Average Infection Peak Magnitude (%population) | Average Peak Time (days) | Peak likelihood |
|---|---|---|---|
| 0% (Lockdown continued as during JNSD) | 2.560421 | 5.436402 | 31% |
| 5% | 2.599640 | 5.510503 | 31% |
| 10% | 2.560560 | 5.408285 | 31% |
| 15% | 2.589800 | 5.527020 | 31% |
| 20% | 3.046286 | 7.439762 | 31% |
| 25% | 3.047949 | 7.723727 | 31% |
| 50%* | 2.720205* | 11.197762* | 27%* |
| 100% (full return to regular activity) | 12.282289 | 18.420751 | 100% |

* Exhibits more than one peak; first peak is described.

Economic reactivation strategies using the modular structure of contact networks limit the spread of the infectious agent. We can observe that a reactivation of up to 25% of the population can be achieved with little deviation from the full lockdown in terms of magnitude, timing, and likelihood of a new peak. In the case of a 50% modular reactivation, subsequent peaks are likely to emerge. It is important to note, however, that the magnitude of these peaks is less than that observed in a full reactivation scenario.

Modular reactivation is thus a better alternative in terms of allowing the incorporation of higher percentages of the population without the risk of massive outbreaks. However, as we can see later on, there are still better alternatives nurtured by the use of a deep knowledge of the topological parameters of the human contact network of Mexico City.

### 4.3 Topology-guided smart selection further mitigates epidemic dynamics

By taking into account not only the global topological modularity of the human contact network of Mexico City but also the way the modules are connected (in the down-scaled version), we have been able to develop an optimized set of reactivation strategies as is presented in Figure 6 and Table 3.

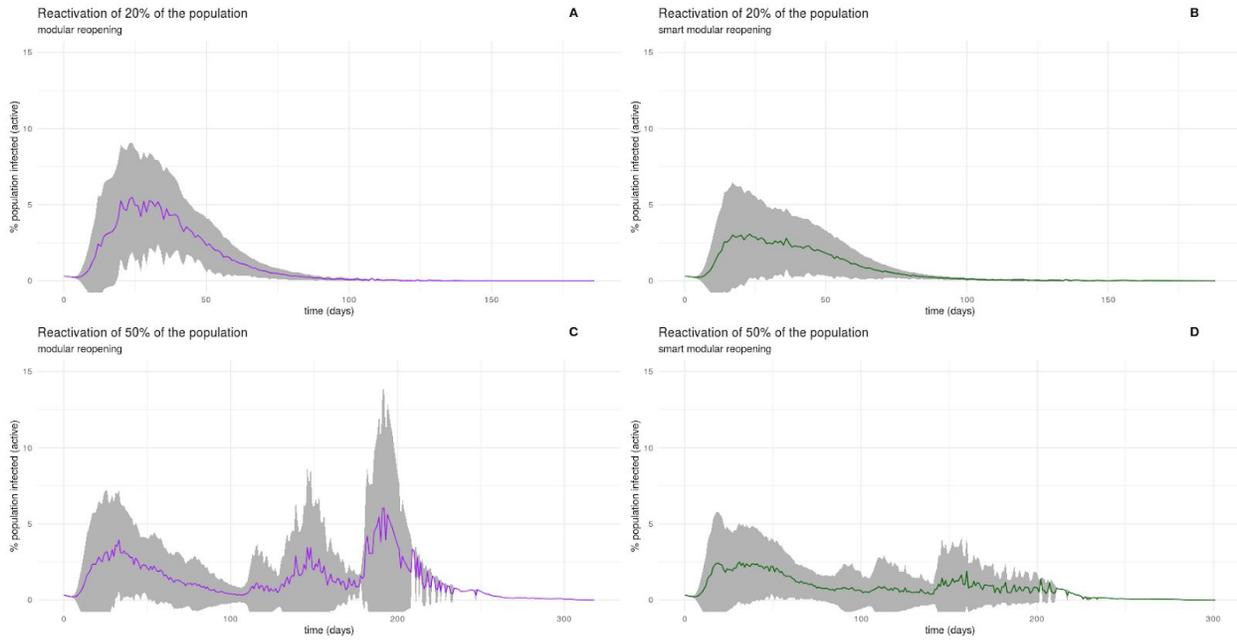

Figure 6: Ensemble visualization of epidemic simulations on networks. The colored line is traced on the average infected population percentage for each time point. The shadowed area represents ±1 standard deviation. Panel A and B represent the dynamics with 20% population reactivation using a modular strategy. For panel A, reactivation is achieved by reactivating 6 modules; whereas the same population fraction was spread in 33 modules for panel B. Panels C and D show the same contrast, this time for a 50% of the population, distributed in 164 and 247 modules respectively. Notice that by spreading the reactivation in several smaller modules, the height of new peaks is reduced.

Table 3: Summary statistics of modular reactivation

| Reactivation scenario | Average Infection Peak Magnitude (%population) | Average Peak Time (days) | Peak likelihood |
|---|---|---|---|
| 20% - 6 modules | 3.046286 | 7.439762 | 31% |
| 20% - 33 modules | 2.496397 | 4.903131 | 30% |
| 50% - 164 modules | 2.720205* | 11.197762* | 27%* |
| 50% - 247 modules | 2.300166* | 8.815707* | 27%* |

\* Exhibits more than one peak; first peak is described.

If the population to be reactivated is spread throughout several (smaller) modules, it is possible to exert more control over the epidemic dynamics. Even in the case of a 50% population reactivation, this "smarter" modular strategy decreases the magnitude of subsequent possible peaks.

Hence, by considering the global and local modularity structure of the contact network and simulating epidemic dynamics compliant with these, we have been able to devise an optimized progressive de-containment strategy.

## 5. Discussion:

**General Discussion**

Epidemic spread of infectious diseases occurs via chains of transmission which are dynamic processes over networks capturing different human behaviors. In the case of COVID-19, the disease is spread through close contact human interactions [Shi, et al 2020], which can be modelled as contact networks. The contact networks

Metropolitan urban environments, in particular large ones such as Mexico City present special challenges for epidemic network modelisation due to their intricate modular structure and size. Computationally efficient methods to scale-down the real (very large) contact networks to manageable yet still descriptive sizes, capturing the relevant aspects of the modular structure of the original networks are needed to perform epidemic dynamic models representative of the real populations.

Using these scaled contact networks, as well as a stochastic dynamics on networks approach, we were able to capture the essentials of epidemic spread in Mexico City. We used the knowledge of said spreading patterns to model de-confinement scenarios to evaluate reactivation strategies after lockdown in Mexico City. Although these networks exhibit a dynamic dimension themselves, using a representative network as a baseline on which different lockdown and reactivation strategies can be modelled provides an efficient tool to explore different scenarios.

Since human contact networks in large urban environments tend to exhibit a modular structure, this invisible compartmentalization is one of the features that shapes the transmission chains of an epidemic phenomenon. By optimizing the modular structure of the re-entrant essential worker population we have been able to propose schemes that allow for a significant percentage of the population to return to public space without leading to massive outbreaks.

In this work, we find that economic reactivation is feasible without necessarily resulting in a new outbreak. However, for this to work, reactivation must occur within contact communities. Optimally, these modules should be small and exhibit minimal connections within them, in order to limit the spread of the disease. A set of behavioral and regulatory actions are needed to tightly constrain this contact dynamics; as the alternative shows that even a small fraction of the population being arbitrarily reactivated leads to an epidemic behavior similar to that of implementing no lockdown at all.

## From the whiteboard to public policy: how to translate

It is well known that implementing health policy from the findings of biomedical research is not a trivial task. A large gap between the volume of research-generated public health knowledge and its application in community settings has been documented. It is quite common that public health scholars find challenging to translate or disseminate their research outcomes to distill them into public policy for use in community settings where it is likely to have positive impacts [Browson, et al 2006].

In clinical fields as is the case of medicine and nursing, a similar void is found between scientific discovery and health policy application. A number of authors have suggested that in order to attain an effective dissemination of scientific findings it is necessary to establish a program with solid community foundations, considering time-efficient approaches, ongoing training, strong organizational values on evidence-based practice. Furthermore, the implementation of a research discovery among government health organizations, clinical practice groups, and the general population is not immediate, but is expected to proceed in stages. The decision to adopt, accept, and utilize an innovation is not an instantaneous act, but more often a process [Giles-Corti, et al 2015].

In the epidemiology of infectious diseases in an open population, such as the case of COVID-19 this situation is actually worsened by the fact that the contact network is not visible in real time. And even if it was, to see it and act on it may involve serious privacy issues. Who do you interact with, how much and how often are questions which result central for the transmission chains to happen. This information may be of a sensible nature or simply, out of reach.

Therefore,there is a need for multiple approaches to properly reach modularization even under such constraints. To this end, modelisation studies may result in enlightenment. The studies just referred have provided us with the following intuitions regarding modularization of the human contact structure in urban environments such as the one in Mexico City:

i) Residential proximity induces communities (people go to the same shops, etc).

ii) Non-public facing workspaces induce communities.

iii) Commuting induces intermodule connections (intermixing in public transportation, recreational outings, etc).

iv) Public facing jobs induce intermodule connections.

Since we know which behaviours involve intra and inter links, we can try to use such correlations in order to limit various at the same time:

**Interleaved reactivation**: different, physically distant neighborhoods every day.

**Delay** the return of workers that require long commutes.

Authorities can **enforce modularity by closing public transportation** needed for intermodular connections; however other public policies will be needed to equilibrate so that workers are not penalized by this. This involves a great degree of **logistic complexity** (beyond the scope of the paper), that must be resolved by the competent authorities.

In order to translate the findings of this study into actionable policy, it is necessary to take into account the aforementioned sources of logistic complexity. In this regard, some measures that may be implemented to encourage modularity reactivation are presented in Table 4.

Table 4. Some examples of policies to encourage/enforce modular reactivation

| Policy | Government actionable | Citizenship actionable | Expected adoption |
|---|---|---|---|
| Encouraging local trade/consumption habits | Suggested | Yes | Heterogeneous |
| Phased workplace reactivation | Suggested or Mandatory | No | Suggested: Heterogeneous<br><br>Mandatory: High |
| Turnover of the workforce by residential address location | Suggested or Mandatory | Yes | Suggested: Heterogeneous<br><br>Mandatory: High |
| Partial transportation shutdown | Complete shutdown or Turned-over schedule of a fraction of public transportation spots/hubs | No | Mandatory:<br><br>High, if combined with complementary policies |

# Concluding remarks

By considering mechanisms for the reactivation of economic activities in Mexico City, we have evaluated the risk that the negative slope of the epidemic curve may be reversed by the effect of the incorporation of a fraction of the population to the public space.

As null models, we ran the epidemic dynamics on the contact network of Mexico City with no mobility restrictions, representing a return to the connectivity prevailing prior to the start of the pandemic. The second null model considered was the network corresponding to the JNSD network which is a subgraph of the Mexico City contact network with just the 25 % of active links, capturing the effect of the reported mobility reduction. In these models, reactivated nodes are randomly distributed on the network, which is a conservative assumption due to lack of further information on the sociodemographic and spatial distribution of the individuals involved in the activities that will be reactivated.

By taking into account the strong need to reactivate economic activities and following the concept of modularity in network theory, we propose the exploration of encapsulation scenarios for new essential workforce reactivation.
A module in a complex network is loosely defined as a set of nodes (individuals) with a higher number of connections among members of the set than with other nodes of the network, i.e. there are more connections within a module than between modules. An important property of modular networks is that dynamic phenomena (such as a random walk or pathogen propagation) tends to remain inside a module for a longer time before spreading outside the module.

By being able to translate the findings of network-based analytics and epidemiological models into actionable public policy measures, it is possible to advance into incorporating predicted risks into the assessment portfolio for reactivation large urban conglomerates such as Mexico City after lockdown in the face of the still ongoing Covid-19 pandemic.